\documentclass[aps,prc,preprint,amsfonts,amsmath,superscriptaddress,floatfix,nofootinbib]{revtex4-2}
\usepackage{graphicx,epsfig,color,amsmath}
\usepackage{soul}
\usepackage{physics}
\usepackage{mathtools}
\usepackage{mathrsfs}
\DeclareMathAlphabet{\mathpzc}{OT1}{pzc}{m}{it}
\usepackage{booktabs}

\newcommand{\beqy}{\begin{eqnarray}}
\newcommand{\eeqy}{\end{eqnarray}}

\newcommand{\D}{\Delta_q}

\newcommand{\mubar}{\bar{\mu}_q}
\newcommand{\muZerobar}{\bar{\mu}_q^{(0)}}
\newcommand{\Vbar}{\bar{\mathbb{V}}_q}

\newcommand{\Tbar}{\bar{T}_q}
\newcommand{\Dbar}{\bar{\Delta}_q}
\newcommand{\DZerobar}{\bar{\Delta}_q^{(0)}}
\newcommand{\Lbar}{\bar{\varepsilon}_{\Lambda}}
\newcommand{\Ex}{\mathbb{E}_x^{(q)}}

\newcommand{\Fermi}{\varepsilon_{Fq}}

\newcommand{\VLq}{\mathbb{V}_{Lq}}
\newcommand{\VLn}{\mathbb{V}_{Ln}}
\newcommand{\Vq}{\mathbb{V}_{q}}
\newcommand{\Vn}{\mathbb{V}_{n}}
\newcommand{\Vcq}{\mathbb{V}_{cq}^{(0)}}

\newcommand{\Yn}{\mathcal{Y}_{n}}
\newcommand{\Yp}{\mathcal{Y}_{p}}

\newcommand{\Euler}{\text{e}}

\newcommand{\QuasipartEnergy}{\mathfrak{E}_{\pmb{k}}^{(q)}}

\newcommand{\mneff}{\dfrac{m_n^{\oplus}}{m_n}}
\newcommand{\mpeff}{\dfrac{m_p^{\oplus}}{m_p}}

\def\apj{Astrophys. J.}  % Astrophysical Journal
\def\apjl{Astrophys. J. Lett.} % Astrophysical Journal, Letters
\def\apss{Astrophys. Space Sci.} 
\def\prl{Phys.~Rev.~Lett.} % Physical Review Letters
\def\prc{Phys.~Rev.~C} % Physical Review C
\def\nphys{Nucl.~Phys.} % Nuclear Physics
\def\mnras{Mon. Not. R. Astron. Soc.}             % Monthly Notices of the RAS
\def\JLowT{J.~Low~Temp.~Phys.} % Journal of Low T physics
\def\RepProgPhys{Rep.~Prog.~Phys.}
\def\japa{J. Astrophys. Astron.}  %Journal of Astrophysics and Astronomy 

\DeclareUnicodeCharacter{2212}{-} %To correct a reccurent error

\begin{document}

\title{Gapless superfluidity in neutron stars: Normal-fluid fraction}

  \author{V. Allard}
 \affiliation{Institute of Astronomy and Astrophysics, Universit\'e Libre de Bruxelles, CP 226, Boulevard du Triomphe, B-1050 Brussels, Belgium}

 \author{N. Chamel}
 \affiliation{Institute of Astronomy and Astrophysics, Universit\'e Libre de Bruxelles, CP 226, Boulevard du Triomphe, B-1050 Brussels, Belgium}

\date{\today}

\begin{abstract}
Our previous investigation within the time-dependent nuclear energy-density functional theory showed 
that the nuclear superfluids contained inside cold neutron stars could become gapless under certain 
circumstances. The absence of a gap in the energy spectrum of quasiparticle excitations leads to a 
specific heat that is comparable to that in the normal phase in sharp contrast with the exponential 
suppression in the BCS phase of type $^1${\it S}$_0$ pairing. Here, we further study gapless superfluidity within the same 
microscopic framework focusing on hydrodynamic properties. In particular, we calculate the mass
fraction transported by the normal fluid of quasiparticle excitations, and we find that it can be 
finite even at zero temperature. We derive approximate analytical formula for arbitrary neutron-proton 
superfluid mixtures. We also present numerical results for neutron stars. Our study 
suggests that the dynamics of neutron stars may be much more complicated than previously thought. 
The realization of gapless superfluidity in neutron stars and its implications are discussed.
\end{abstract}

\maketitle

\section{Introduction}

Superfluid helium exhibits hydrodynamical phenomena, such as the fountain effect~\cite{allenjones1938}, not seen in ordinary fluids. These phenomena are the manifestations of the interpenetration of two distinct dynamical components: a normal viscous fluid with mass density $\rho_{\rm N}$ and a superfluid with mass density $\rho_{\rm S}=\rho-\rho_{\rm N}$, where $\rho$ is the mass density of helium. Both components transport mass but with different velocities. Unlike the normal fluid, the superfluid carries no entropy and flows without resistance. This two-fluid picture was originally proposed by Tisza~\cite{Tisza1938} based on the concept of Bose-Einstein condensation put forward by London~\cite{london1938,london1938b}; the normal component was then identified with non-condensed atoms. This model was later reformulated by Landau in terms of quasiparticle excitations in quantum fluids~\cite{Landau1941} (see also Ref.~\cite{khalatnikov1989introduction}). Since no excitation is present in the ground state, the normal fluid only exists at finite temperature $T$. The mass fraction $\rho_{\rm N}/\rho$ increases with $T$ as more and more quasiparticles are excited, and reaches unity at the critical temperature for which superfluidity is destroyed. 

Soon after Bardeen-Cooper-Schrieffer (BCS) demonstrated in the context of superconductivity that fermions could also condense by forming pairs~\cite{Bardeen1957}, it was realized that the interior of neutron stars, the compact remnants of gravitational core-collapse supernova explosions, could be superfluid~\cite{migdal1959,ginzburg1965}. This prediction found strong support from the first 
pulsar observations~\cite{baym1969}. Similarly to superfluid helium, cold neutron stars 
are generally described as a two-fluid mixture consisting of a neutron superfluid and a proton superconductor; the conglomerate of rigidly coupled charged particles (nuclei in the crust plus leptons in the core) are essentially co-moving with the latter (see, e.g., Ref.~\cite{andersson2021} for a recent review). 
Relative flows between these two fluids arising from the acceleration or deceleration of charged particles due to accretion from a stellar companion or electromagnetic braking, can be sustained by the pinning of neutron quantised vortices. 
Global readjustments in the rotational motions of the two fluids induced by the unpinning of vortices are expected to be at the origin 
of the largest observed pulsar frequency glitches
(see, e.g., Refs.~\cite{antonopoulou2022,zhou2022} for recent reviews). 
The multifluid dynamics in superfluid neutron stars can leave its imprint on various other astrophysical phenomena, such as oscillations and gravitational-wave emission~\cite{andersson2021gw}.

Despite the absence of viscous drag, the two fluids are still weakly coupled by non-dissipative 
mutual entrainment effects caused by nuclear interactions~\cite{sauls1989superfluidity}, of the same kind as the ones discussed earlier by Andreev and Bashkin~\cite{andreev1976} in the context of superfluid $^{4}$He-$^3$He mixtures. The magnetization of neutron vortices induced by the circulation of 
entrained protons could also give rise to some effective friction between the two fluids~\cite{alpar1984}. The entrainment couplings are therefore 
key microscopic parameters for hydrodynamical simulations of superfluid neutron stars, and can 
be expressed in terms of generalized superfluid densities. In a series of 
papers~\cite{ChamelAllard2019,ChamelAllard2020,ChamelAllard2021}, we have explicitly calculated these densities in the outer core of a neutron star (where nucleons form $^1${\it S}$_0$ type of Cooper pairs) for arbitrary temperatures and 
currents within the time-dependent nuclear energy-density functional theory (see, e.g. Ref.~\cite{schunck2019}) by solving exactly the self-consistent time-dependent Hartree–Fock–Bogoliubov (TDHFB) equations. These densities were previously calculated within the Fermi liquid theory (see Ref.~\cite{leinson2018} and references therein). Results were also obtained using relativistic mean-field models, though not for arbitrary currents and temperatures (see Ref.~\cite{sourie2016} and references therein). Discussions about earlier calculations can be found in Refs.~\cite{ChamelAllard2019,ChamelAllard2020}. 

More recently, we have shown that the nuclear energy-density functional theory (therefore also the Fermi liquid theory since it can be derived as an approximation in the limit of small temperatures and small velocities compared to their Fermi counterparts; see Section 2.5 of Ref.~\cite{ChamelAllard2021}) leads to the existence of a nucleon superfluid state in which the energy spectrum of quasiparticle excitations becomes gapless under certain circumstances~\cite{AllardChamel2023PartI}. In our previous paper, we focused on the associated thermal properties. 
Here, we investigate more closely the hydrodynamic 
properties of neutron-proton superfluid mixtures in this gapless regime. In particular, we calculate in Section~\ref{sec:superfluid-mixtures} the mass density associated with the normal fluid of quasiparticle excitations for cold homogeneous neutron-proton superfluid mixtures. Applications to neutron stars are 
presented and discussed in Section~\ref{sec:applications}. The realization of gapless superfluidity in 
neutron stars is discussed in Section~\ref{sec:conclusions}. 

\section{Dynamics of cold homogeneous neutron-proton superfluid mixtures}
\label{sec:superfluid-mixtures}

Throughout this paper, we consider $^1${\it S}$_0$ superfluidity only, as predicted to occur in the inner crust and in the outer core of neutron stars. Neutrons may form $^3${\it PF}$_2$ pairs at higher densities with critical temperatures of order 0.1 MeV or less (see, e.g., Ref.~\cite{sedrakian2019} for a review), however the existence of such a superfluid phase in neutron stars remains uncertain~\cite{yasui2020}. It has been recently found that neutrons could pair in the $^3P_0$ channel instead~\cite{Krotscheck2023,krotscheck2023b}.

\subsection{Quasiparticle excitations and gapless superfluid phase}

Within the nuclear energy-density functional theory based on Skyrme effective interactions~\cite{bender03} and extended versions allowing for terms that are both momentum- and density-dependent~\cite{chamelgoriely2009}, the energy of a quasiparticle excitation with momentum $\hbar \pmb{k}$ ($\hbar$ is the Planck-Dirac constant) in an homogeneous neutron-proton superfluid mixture with stationary flows in the normal fluid rest frame at temperature $T$, is given by~\cite{ChamelAllard2020} ($q=n,p$ for neutrons and protons respectively)
\begin{align}\label{eq:QuasiparticleEnergy}
    \mathfrak{E}_{\pmb{k}}^{(q)}=\hbar\pmb{k}\cdot\pmb{\mathbb{V}_q} + \sqrt{\varepsilon^{(q)2}_{\pmb{k}} +  \Delta_q^2}\, ,
\end{align}
with 
\begin{align}\label{eq:varepsilon}
\varepsilon^{(q)}_{\pmb{k}} =\frac{\hbar^2 k^2}{2m^{\oplus}_q}  - \mu_q \, ,
\end{align}
and we have introduced the effective ``superfluid velocity''
\begin{align}\label{eq:EffectiveSuperfluidVelocity}
    \pmb{\mathbb{V}_q}\equiv \frac{m_q}{m_q^{\oplus}}\pmb{V_q}+\frac{\pmb{I_q}}{\hbar}\, ,  
\end{align}
where $\pmb{V_q}$ is the usual superfluid velocity. 
Here, $m_q$ denotes the relevant nucleon mass, $\mu_q$ is the reduced chemical potential, $m_q^\oplus$ is the effective mass, $\pmb{I_q}$ the vector mean-field potential. The quantity $\D$ is related to the complex order parameter of the superfluid phase given by 
\beqy
\Psi_q(\pmb{r}) = \frac{\D}{v^{\pi q}}\exp\left(\frac{2i m_q \pmb{V_q}\cdot \pmb{r}}{\hbar}\right)\; .  
\eeqy
It is obtained from the self-consistent equation
\begin{align}
\label{eq:GapEquation}
    \D(T,\pmb{\Vq}) = -\frac{1}{2V} v^{\pi q} \sum_{\pmb{k}} \frac{\D(T,\pmb{\Vq})}{\sqrt{\varepsilon_{\pmb{k}}^{(q)2} +\D(T,\pmb{\Vq})^2}}\tanh\left(\frac{\beta}{2}\QuasipartEnergy\right)\, ,
\end{align}
where $V$ is the normalization volume, 
$\beta = \left(k_\text{B}T\right)^{-1}$ ($k_\text{B}$ being the Boltzmann constant and $T$, the temperature) and $v^{\pi q}<0$ denotes the pairing strength. It is understood that the summation must be regularized to remove ultraviolet divergences (see, e.g.,  Refs.~\cite{bulgac2002,schunck2019} for discussions). 
Equation~\eqref{eq:GapEquation} must be solved together with the particle number conservation
\begin{align}
\label{eq:DensityHomogeneous}
n_q=
\frac{1}{V}\sum_{\pmb{k}}\left[1-\frac{\varepsilon^{(q)}_{\pmb{k}}}{\sqrt{\varepsilon^{(q)2}_{\pmb{k}} +  \D^2}}\tanh\left(\frac{\beta}{2}\QuasipartEnergy\right)\right]\, , 
\end{align}
with $n_q$ being the nucleon number density. See Ref.~\cite{ChamelAllard2020} for more details. 

In the absence of superflow ($\Vq=0$), $\D$ characterizes the gap in the quasiparticle energy spectrum. However, we have recently shown that this is no longer the case for finite $\Vq$ at zero temperature: the quasiparticle energy gap shrinks with increasing $\Vq$ and vanishes at Landau's velocity $\VLq$ (see Eq.~\eqref{eq:Landau-effective-velocity} below) while $\D$ remains unchanged. 
For higher effective superfluid velocities, $\D$ decreases and drops to zero at some critical velocity $\Vcq$ beyond which superfluidity disappears. In the gapless regime $\VLq<\mathbb{V}_q < \Vcq $, a normal fluid made of quasiparticle excitations therefore coexists with the superfluids even though $T=0$~\cite{AllardChamel2023PartI}. According to the third law of thermodynamics, the normal fluid at $T=0$ cannot carry any entropy. This can 
also be seen from the well-known expression for the entropy density at arbitrary temperature $T$, obtained within the HFB approach (see, e.g., Ref.~\cite{blaizot1986})
\beqy \label{eq:Entropy}
s_q=-\frac{2 k_{\rm B}}{V} \sum_{\pmb{k}} \left[f^{(q)}_{\pmb{k}}\log f^{(q)}_{\pmb{k}} +(1-f^{(q)}_{\pmb{k}})\log(1-f^{(q)}_{\pmb{k}})\right]\, ,
\eeqy 
where $f^{(q)}_{\pmb{k}}$ denotes the quasiparticle distribution function at temperature $T$, given by
\begin{equation}\label{eq:QuasiparticleDistribution}
    f^{(q)}_{\pmb{k}} = \left[1+\exp\left(\beta \mathfrak{E}_{\pmb{k}}^{(q)}\right)\right]^{-1}=\frac{1}{2}\left[1-\tanh\left(\frac{\beta}{2} \mathfrak{E}_{\pmb{k}}^{(q)}\right)\right]\, .
\end{equation}
In the limit $T\rightarrow 0$, $f^{(q)}_{\pmb{k}}$ vanishes if $\mathbb{V}_q < \mathbb{V}_{Lq}$ since $\mathfrak{E}_{\pmb{k}}^{(q)}> 0$ for all wave vectors $\pmb{k}$. For effective superfluid velocities $\mathbb{V}_{Lq}<\mathbb{V}_q < \mathbb{V}^{(0)}_{cq}$, there exist quasiparticle states such that $\mathfrak{E}_{\pmb{k}}^{(q)}< 0$  
implying $f^{(q)}_{\pmb{k}}=1$. In the gapless regime, $f^{(q)}_{\pmb{k}}$ is therefore equal to either 0 or 1. It follows from Eq.~\eqref{eq:Entropy} that $s_q=0$ still holds, as expected. 

\subsection{Normal-fluid fraction at zero temperature}

Although the normal fluid at $T=0$ does not carry entropy, it does carry mass and momentum. 
Let us recall that at any temperature the mass current $\pmb{\rho_q}$ of one nucleon species is generally expressible as a combination of the superfluid velocities $\pmb{V_q}$ of both species and of the normal-fluid velocity $\pmb{v_N}$,  as~\cite{gusakov2005} 
\beqy\label{eq:entrainment1}
\pmb{\rho_n} &=& \rho_n^{\rm (N)}\pmb{v_{N}} + \rho_{nn} \pmb{V_{n}}+\rho_{np} \pmb{V_{p}}\, , \\  
\label{eq:entrainment2}
\pmb{\rho_p} &=& \rho_p^{\rm (N)} \pmb{v_{N}} + \rho_{pn} \pmb{V_{n}}+\rho_{pp} \pmb{V_{p}}\, ,
\eeqy
where the (symmetric) entrainment matrix $\rho_{qq'}$ ($q,q'=n, p$) characterizes the strength of the entrainment coupling and is the generalization of the concept of superfluid density to mixtures. The normal nucleon densities (i.e. the mass density carried by the excitations of nucleon species $q$) are related to the entrainment matrix through the following relations~\cite{gusakov2005} : 
\beqy\label{eq:NormalDensity1}
\rho_n^{\rm (N)}&=&\rho_n - \rho_{nn}-\rho_{np}\, , \\
\label{eq:NormalDensity2}
\rho_p^{\rm (N)}&=&\rho_p - \rho_{pp}-\rho_{pn}\, ,
\eeqy
$\rho_q=m_q n_q$ denoting the mass density of nucleon of charge type $q$. 
Therefore the total mass density carried by the normal fluid is given by 
\beqy\label{eq:normal-fluid-fraction}
\rho_{\rm N}= \rho_n^{\rm (N)} + \rho_p^{\rm (N)}=\rho-\rho_{nn}-\rho_{pp}-2\rho_{np}\, , 
\eeqy 
and we have introduced the total mass density $\rho=\rho_p + \rho_n$. It can be immediately seen that at $T=0$ the normal fluid does not exist ($\rho_q^{\rm (N)}=0$) if $\mathbb{V}_q<\mathbb{V}_{Lq}$ since in this case the following identities hold (see, e.g. Ref.~\cite{ChamelAllard2019}): 
\beqy 
\rho_{nn}+\rho_{np}=\rho_n \\ 
\rho_{pp}+\rho_{pn}=\rho_p \, .
\eeqy 
It is only in the gapless phase $\mathbb{V}_{Lq}<\mathbb{V}_q < \mathbb{V}^{(0)}_{cq} $ that we have $\rho_q^{\rm (N)}\neq 0$ at $T=0$. 

As discussed in Refs.~\cite{ChamelAllard2020,ChamelAllard2021}, the superfluid velocities $\pmb{V_q}$ are actually not true velocities but characterize momenta per nucleon. It thus follows from Galilean invariance that  the normal fluid carries a momentum density defined by
\beqy
\pmb{\Pi_{\rm N}}\equiv\pmb{\rho_n} +\pmb{\rho_p} - \rho_{n} \pmb{V_{n}}-\rho_{p} \pmb{V_{p}}\, .
\eeqy
Using Eqs.~\eqref{eq:entrainment1}, \eqref{eq:entrainment2}, and \eqref{eq:normal-fluid-fraction}, this momentum density is given by
\beqy
\pmb{\Pi_{\rm N}}=\rho_N \pmb{v_N} - \rho_n^{\rm (N)} \pmb{V_n} - \rho_p^{\rm (N)} \pmb{V_p}\, .
\eeqy

The exact expressions of the entrainment matrix for arbitrary temperatures and effective superfluid velocities within the TDHFB theory have been obtained in Refs.~\cite{ChamelAllard2019,ChamelAllard2020} and read:   
\begin{align}\label{eq:EntrainmentMatrix}
    \rho_{qq'}=\rho_q\left(1-\mathcal{Y}_q\right)\left(\frac{m_q}{m_q^{\oplus}}\delta_{qq'} + \frac{\mathcal{I}_{qq'}}{\hbar}\right)
\end{align}
with
\begin{align}\label{eq:Inn-def}
\mathcal{I}_{nn}&=\frac{2}{\hbar}\rho_{n}  \left(1-\mathcal{Y}_{n}\right)\Theta\left[C_1^\tau\left(\frac{8}{\hbar^2}C_0^\tau m_p^{\oplus}n_p\mathcal{Y}_p-1\right)-C_0^\tau\right]\, ,
\end{align}

\begin{align}\label{eq:Ipp-def}
\mathcal{I}_{pp}&=\frac{2}{\hbar}\rho_{p}  \left(1-\mathcal{Y}_{p}\right)\Theta\left[C_1^\tau\left(\frac{8}{\hbar^2}C_0^\tau m_n^{\oplus}n_n\mathcal{Y}_n-1\right)-C_0^\tau\right]\, ,
\end{align}

\begin{align}\label{eq:Inp-def}
\mathcal{I}_{np}&=\frac{2}{\hbar}\rho_{p}  \left(1-\mathcal{Y}_{p}\right)\Theta\left(C_1^\tau-C_0^\tau\right)\, ,
\end{align}

\begin{align}\label{eq:Ipn-def}
\mathcal{I}_{pn}&=\frac{2}{\hbar}\rho_{n}  \left(1-\mathcal{Y}_{n}\right)\Theta\left(C_1^\tau-C_0^\tau\right)\, ,
\end{align} 

\begin{align}\label{eq:Theta-def}
\Theta &\equiv\Biggl[ 1- \frac{2}{\hbar^2}\left(C_0^\tau +C_1^\tau\right)\left(m_n^{\oplus}n_{n}\mathcal{Y}_{n} + m_p^{\oplus}n_{p}\mathcal{Y}_{p} \right)\nonumber \\
&\qquad\;\;  +\left(\frac{4}{\hbar^2}\right)^2 C_0^\tau C_1^\tau m_n^{\oplus}n_{n}m_p^{\oplus}n_{p}\mathcal{Y}_{n} \mathcal{Y}_{p} \Biggr]^{-1}\, .
\end{align}
Here $C_0^\tau$ and $C_1^\tau$ are coupling coefficients associated with isoscalar and isovector dynamical terms in the nuclear energy-density functional (see Appendix B of Ref.~\cite{ChamelAllard2019}). They are related to the effective masses through the following equation:
\beqy\label{eq:def-Bq}
\frac{m_q}{m_q^\oplus} =1+ \frac{2 \rho}{\hbar^2}(C_0^\tau-C_1^\tau)  + \frac{4 \rho_q}{\hbar^2} C_1^\tau
\, .
\eeqy

The function $\mathcal{Y}_q$, defined by 
\begin{align}\label{eq:YqFunction}
\mathcal{Y}_q(T,\pmb{\mathbb{V}_q})\equiv \frac{\hbar}{m_q^{\oplus}n_q \mathbb{V}_q^2}\frac{1}{V}\sum_{\pmb{k}} \pmb{k}\cdot\pmb{\mathbb{V}_q}\tanh\left(\frac{\beta}{2}\mathfrak{E}^{(q)}_{\pmb{k}}\right)\, ,
\end{align}
has a simple physical interpretation in cases for which the effective masses coincide 
with the bare masses\footnote{According to Eq.~\eqref{eq:def-Bq}, $m_n^\oplus=m_n$ and $m_p^\oplus=m_p$ for nuclear energy-density functionals such that $C_0^\tau=0=C_1^\tau$.}: it can then be identified with the normal-fluid fraction
associated with quasiparticle excitations of the nucleon species $q$. 
Indeed, setting $m_n^\oplus=m_n$ and $m_p^\oplus=m_p$, the normal-fluid densities, which can be obtained from the definitions~\eqref{eq:NormalDensity1},~\eqref{eq:NormalDensity2} and~\eqref{eq:normal-fluid-fraction} using Eqs.~\eqref{eq:EntrainmentMatrix}-\eqref{eq:Theta-def}, reduce to
\beqy\label{eq:normal-density-no-effmass}
&\rho_n^{\rm (N)}=\rho_n \mathcal{Y}_n\; , \qquad\qquad \rho_p^{\rm (N)}=\rho_p\mathcal{Y}_p\;, \\
&\rho_{\rm N}=\rho_n \mathcal{Y}_n+\rho_p \mathcal{Y}_p \, .
\eeqy
In general, however, the normal-fluid densities take a much more complicated form and the physical interpretation of the function $\mathcal{Y}_q$ is less straightforward.

In the limit of pure neutron matter, 
the normal-fluid fraction is expressible as 
\beqy \label{eq:normal-fraction-approx2}
\frac{\rho^{\rm (N)}_n}{\rho_n}=\dfrac{m_n^\oplus}{m_n}\frac{\mathcal{Y}_n}{1+\left(\dfrac{m_n^\oplus}{m_n}-1\right)\mathcal{Y}_n}\, .
\eeqy 
This expression is valid for arbitrary temperature and neutron effective superfluid velocity (note that $\pmb{\mathbb{V}_n}=\pmb{V_n}$ in this limiting case). 
In the absence of current, this expression reduces to that obtained by Leggett~\cite{leggett1965} 
within his extension of Landau’s theory to what he called ``superfluid Fermi liquids'' in the weak-coupling approximation (see Appendix). 

For arbitrary composition, the normal-fluid densities can be expressed in terms 
of $\rho_n$, $\rho_p$, $m_n^\oplus$, $m_p^\oplus$, $\mathcal{Y}_n$ and $\mathcal{Y}_p$. Their variations with respect to the effective superfluid velocity $\pmb{\Vq}$ and the temperature $T$ are therefore  entirely contained  in the functions $\mathcal{Y}_q(T,\pmb{\Vq})$. An explicit expression of this function at $T=0$ in the gapless superfluid phase is derived in the next subsection.

\subsection{Analytic expression of the function $\mathcal{Y}_q$ in the gapless superfluid phase}

We take the continuum limit of  Eq.~\eqref{eq:YqFunction}, i.e. we replace discrete summations over wave vectors by integrations as follows: 
\beqy\label{eq:continuum}
\frac{1}{V} \sum_{\pmb{k}} \dotsi \rightarrow   \int \frac{{\rm d}^3\pmb{k}}{(2\pi)^3} \dotsi =\int \frac{{\rm d}\Omega_{\pmb{k}}}{4 \pi}\int {\rm d}\varepsilon\, \mathcal{D}_q(\varepsilon) \dotsi
\eeqy 
with $\Omega_{\pmb{k}}$ the solid angle in $\pmb{k}$-space, $\mathcal{D}_q(\varepsilon)$ the density of single-particle states per spin given by 
\beqy\label{eq:DoS}
\mathcal{D}_q(\varepsilon)\equiv \int \frac{{\rm d}^3\pmb{k}}{(2\pi)^3} \delta(\varepsilon-\varepsilon_{\pmb{k}}^{(q)})=\frac{m_q^\oplus}{2 \pi^2 \hbar^3}\sqrt{2 m_q^\oplus(\varepsilon+\mu_q)}\, ,
\eeqy 
and we have made use of Eq.~\eqref{eq:varepsilon}. 
Introducing the dimensionless ratios 
\beqy\label{eq:BarQuantities}
\Tbar = \frac{T}{T_{Fq}}\, , \quad \mubar=\frac{\mu_q}{\Fermi}\, ,  \quad  \Dbar=\frac{\D}{\Fermi}\, , \quad \Vbar = \frac{\mathbb{V}_q}{V_{Fq}}\, , 
\eeqy
with the Fermi energy $\Fermi=\hbar^2k_{Fq}^2/(2m_q^{\oplus})$, the Fermi temperature $T_{Fq}=\Fermi/k_\text{B}$ and the Fermi velocity $V_{Fq}=\hbar k_{Fq}/m_q^{\oplus}$ (recalling that the Fermi wave-number is given by $k_{Fq}=(3\pi^2 n_q)^{1/3}$), changing variables and integrating over solid angles (in $\pmb{k}$-space) yield~\cite{ChamelAllard2021} 
\begin{align}\label{eq:Yq}
\mathcal{Y}_q(T,\mathbb{V}_q)=&\frac{3}{8}\frac{\Tbar}{\Vbar^2}\int_0^{+\infty}\text{d}x\;\sqrt{x} \log\left\{\left[1+\text{e}^{-\left(\Ex - 2\Vbar\sqrt{x}\right)/\Tbar}\right]\left[1+\text{e}^{-\left(\Ex + 2\Vbar\sqrt{x}\right)/\Tbar}\right]\right\}\nonumber\\
& +\frac{3}{16}\frac{\Tbar^2}{\Vbar^3}\int_0^{+\infty}\text{d}x\; \left\{\text{Li}_2\left[-\text{e}^{-\left(\Ex - 2\Vbar\sqrt{x}\right)/\Tbar}\right]-\text{Li}_2\left[-\text{e}^{-\left(\Ex + 2\Vbar\sqrt{x}\right)/\Tbar}\right]\right\} \, .
\end{align}
Here Li$_2$ denotes the dilogarithm and 
\begin{align}
    \Ex\equiv \sqrt{\left(x-\mubar\right)^2+\Dbar^2}\, .
\end{align}

If the superfluid velocity is small enough such that $\Ex > 2 \Vbar \sqrt{x}$  for all $x$, i.e. if $\mathbb{V}_q<\mathbb{V}_{Lq}$ as shown in our previous paper, $\mathcal{Y}_q=0$ at $T=0$ independently of $\mathbb{V}_q$:
\beqy \label{eq:YqTZeroSubgapless}
\mathcal{Y}_q(T=0,\mathbb{V}_q<\mathbb{V}_{Lq})=0\, .
\eeqy 

If $\mathbb{V}_q>\mathbb{V}_{Lq}$ we have seen that $\Ex < 2 \Vbar \sqrt{x}$ for $x_-<x<x_+$ with 
\beqy \label{eq:xpm}
x_\pm = \mubar +2\Vbar^2 \pm 2 \sqrt{\mubar \Vbar^2 +\Vbar^4-\frac{1}{4}\Dbar^2} \, .
\eeqy 
Taking the limit $T\rightarrow 0$ and using the expansion 
\beqy 
\mathrm{Li}_2 (u) \approx -\frac{1}{2}\left[ \log(-u)^2 + \frac{\pi^2}{3} \right]
\eeqy 
in the limit $u\rightarrow -\infty$, Eq.~\eqref{eq:Yq} can be integrated analytically\footnote{Note that the contribution to the integrals in the right-hand side of Eq~\eqref{eq:Yq} from $x$ values outside the $[x_{-};x_{+}]$ interval vanishes identically.} leading to 
\begin{align}
\mathcal{Y}_q(T=0,\mathbb{V}_q>\mathbb{V}_{Lq}) =& \frac{3}{16 \Vbar}(x_+^2-x_-^2) - \frac{1}{32\Vbar^3}\left[ (x_+-\mubar)^3-(x_--\mubar)^3\right]\nonumber\\ 
&-\frac{3\Dbar^2}{32\Vbar^3}(x_+-x_-) \, .
\end{align}
Substituting Eq.~\eqref{eq:xpm} and after some simplifications, we finally obtain
\beqy \label{eq:YqTZero}
\mathcal{Y}_q(T=0, \VLq \leq \mathbb{V}_q\leq \Vcq) 
=\left(\mubar+\Vbar^2-\frac{1}{4}\frac{\Dbar^2}{\Vbar^2}\right)^{3/2} \, .
\eeqy

\subsection{Order parameter and chemical potential}

In Eq.~\eqref{eq:YqTZero}, $\mubar$ and $\Dbar$ depend on $\Vq$, as well as on the nucleon densities $n_n$ 
and $n_p$. They are determined by the solutions of  Eqs.~\eqref{eq:GapEquation} and \eqref{eq:DensityHomogeneous}, which in the continuum limit become 
respectively~\cite{ChamelAllard2021}
\beqy\label{eq:DimensionlessGapEquation}
1&=& 
-\frac{1}{2}\left(\frac{k_{Fq}m^{\oplus}_q}{2\pi^2 \hbar^2}\right) v^{\pi q} \frac{\Tbar}{2\Vbar}\displaystyle \int_0^{\mubar + \Lbar}\frac{\text{d}x}{\Ex} \nonumber \\
&&\qquad\times \log\left[\cosh\left(\frac{\Ex}{2\Tbar} +\frac{\Vbar}{\Tbar}\sqrt{x}\right) \sech\left(\frac{\Ex}{2\Tbar} -\frac{\Vbar}{\Tbar}\sqrt{x}\right)  \right]\, ,
\eeqy
\beqy\label{eq:DimensionlessChemicalPotentialEquation}
\frac{4}{3}&=&\int_0^{+\infty}\text{d}x\;\left\{\sqrt{x}-\frac{\Tbar}{\Vbar}\frac{x-\mubar}{2\Ex} \right. \nonumber \\ 
&&\qquad\left.\times \log\left[\cosh\left(\frac{\Ex}{2\Tbar} +\frac{\Vbar}{\Tbar}\sqrt{x}\right) \sech\left(\frac{\Ex}{2\Tbar} -\frac{\Vbar}{\Tbar}\sqrt{x}\right)  \right]\right\}\, .
\eeqy
At low temperatures $\Tbar\ll 1$, the logarithm in the integrals can admit two asymptotic limits depending on the $x$ values. For $\Ex < 2\Vbar\sqrt{x}$ (equivalent to $x_{-}<x<x_{+}$),
\beqy\label{eq:Lim1}
\log\left[\cosh\left(\frac{\Ex}{2\Tbar} +\frac{\Vbar}{\Tbar}\sqrt{x}\right) \sech\left(\frac{\Ex}{2\Tbar} -\frac{\Vbar}{\Tbar}\sqrt{x}\right)\right]\approx \frac{\Ex}{\Tbar} \, .
\eeqy
The other case $\Ex > 2\Vbar\sqrt{x}$, corresponding to $x$ values below $x_{-}$ or above $x_{+}$, leads to 
\beqy\label{eq:Lim2}
\log\left[\cosh\left(\frac{\Ex}{2\Tbar} +\frac{\Vbar}{\Tbar}\sqrt{x}\right) \sech\left(\frac{\Ex}{2\Tbar} -\frac{\Vbar}{\Tbar}\sqrt{x}\right)\right]\approx 2\frac{\Vbar}{\Tbar}\Ex \, . 
\eeqy
Using these asymptotic forms and taking the limit $\Tbar\rightarrow 0$, Eq.~\eqref{eq:DimensionlessGapEquation} reduces to
\beqy\label{eq:GapZeroT}
1=-\frac{1}{2}\left(\frac{k_{Fq}m^{\oplus}_q}{2\pi^2 \hbar^2}\right) v^{\pi q}\left[ \int_0^{x_{-}}\text{d}x\; \frac{\sqrt{x}}{\Ex}  +\int_{x_{+}}^{+\infty}\text{d}x\; \frac{\sqrt{x}}{\Ex} - \frac{1}{2\Vbar}\left(x_{+}-x_{-}\right)\right]\, ,
\eeqy
while Eq.~\eqref{eq:DimensionlessChemicalPotentialEquation} becomes
\beqy\label{eq:MuZeroT}
\frac{4}{3}&=&\int_0^{x_{-}}\text{d}x\; \sqrt{x}\left(1-\frac{x-\mubar}{\Ex}\right)+\int_{x_{+}}^{+\infty}\text{d}x\; \sqrt{x}\left(1-\frac{x-\mubar}{\Ex}\right) \nonumber \\ 
&&\qquad + \frac{2}{3}\left(x_{+}^{3/2}-x_{-}^{3/2}\right)+\frac{1}{2\Vbar}\left[x_{+}\left(\mubar-\frac{x_{+}}{2}\right)-x_{-}\left(\mubar-\frac{x_{-}}{2}\right)\right]\, .
\eeqy
Let us recall that $x_{+}$ and $x_{-}$ depend on $\Dbar$ and $\mubar$, and are given by Eq.~\eqref{eq:xpm}. Solving Eqs.~\eqref{eq:GapZeroT} and~\eqref{eq:MuZeroT} yields $\Delta_q (T=0,\Vq\geq \VLq)$ and $\mu_q(T=0,\Vq\geq \VLq)$. 

Note that for $\Vq\leq\VLq$, we have $\Ex \geq 2 \Vbar \sqrt{x}$ for all $x$. In such case, the logarithm can be 
approximated by Eq.~\eqref{eq:Lim2} so that Eqs.~\eqref{eq:DimensionlessGapEquation} and ~\eqref{eq:DimensionlessChemicalPotentialEquation} do not depend on $\Vq$. It follows that 
\beqy 
\Delta_q (T=0,\Vq\leq\VLq)=\Delta_q(T=0,\Vq=0)\equiv\Delta_q^{(0)}\, , 
\eeqy 
and 
\beqy 
\mu_q (T=0,\Vq\leq\VLq)=\mu_q(T=0,\Vq=0)\equiv\mu_q^{(0)}\, ,
\eeqy 
as shown in Sec.III.B. in our previous article~\cite{AllardChamel2023PartI}. 

\subsection{Effective superfluid velocities delimiting the gapless phase}

Landau's effective superfluid velocity is explicitly given by~\cite{AllardChamel2023PartI} 
\beqy\label{eq:Landau-effective-velocity}
\VLq = V_{Fq}\sqrt{\frac{\muZerobar}{2}\Biggl[ \sqrt{1+\left(\frac{\DZerobar}{\muZerobar}\right)^2}-1\Biggr] } \, .
\eeqy 
Let us recall that $\muZerobar\equiv \mubar(T=0,\Vq\leq \VLq)$ and $\DZerobar\equiv \Dbar (T=0,\Vq\leq\VLq)$ are independent of $\Vq$ for $\Vq\leq\VLq$, and can be calculated here in the absence of superflows.

The critical velocity $\Vcq$ is determined by the solution of the 
following equation 
\begin{align}\label{eq:Vcq}
&\mathcal{I}_q\biggl[\mu_q(T=0,\Vq=0);\Delta_q(T=0,\Vq=0)\biggr] \nonumber \\ 
&=2 \sqrt{1 + \Bar{\mathbb{V}}_{cq}^{(0)2}}   -2 \sqrt{x^c_-} + 2 \, \mathrm{arctanh}\sqrt{x^c_-} 
+2 (\sqrt{1+\Lbar} - \sqrt{x^c_+}) \nonumber \\ 
&\qquad +2  \left( \mathrm{arcoth} \sqrt{x^c_+}-\mathrm{arcoth} \sqrt{1+\Lbar} \right)
\, ,
\end{align}
where 
\beqy\label{eq:Integral1}
\mathcal{I}_q\biggl[\mu_q(T,\Vq);\Delta_q(T,\Vq)\biggr] = \int_0^{\mubar+\Lbar} \text{d}x \sqrt{\frac{x}{
\left(x-\mubar\right)^2+\Dbar^2}}\, ,
\eeqy
and $\Lbar$ is the energy cutoff in units of the Fermi energy that must be introduced to regularize Eq.~\eqref{eq:DensityHomogeneous}. Equation~\eqref{eq:Vcq} must be solved together with Eq.~\eqref{eq:MuZeroT} where $\Dbar=0$, $\mubar = \Bar{\mu}_{cq}$ and 
$x_\pm=x_\pm^{c} $ are evaluated at $\Vcq$: 
\begin{align} \label{eq:xZeroTVc}
x_\pm^{c} = \Bar{\mu}_{cq} + 2 \Bar{\mathbb{V}}_{cq}^{(0)} \left( \Bar{\mathbb{V}}_{cq}^{(0)} \pm \sqrt{ \Bar{\mu}_{cq} + \Bar{\mathbb{V}}_{cq}^{(0)2}}\right)\; . 
\end{align}
Equation~\eqref{eq:MuZeroT} thus becomes 
\begin{align}\label{eq:MuZeroTVc}
1=\frac{1}{2}\left(x_{+}^{c\; 2/3}+x_{-}^{c\; 2/3}\right)+\frac{3}{8\Bar{\mathbb{V}}_{cq}^{(0)}}\left[ x^{c}_{+}\left(\bar{\mu}_{cq}-\frac{x^{c}_{+}}{2}\right)-x^{c}_{-}\left(\bar{\mu}_{cq}-\frac{x^{c}_{-}}{2}\right)\right]\, .
\end{align}

\subsection{Weak-coupling approximation}

Let us emphasize that no approximation has been made so far. 
In the weak-coupling regime $\mubar\approx 1$ and $\Vbar \ll 1$, Eq.~\eqref{eq:YqTZero} can be expressed as 
\beqy \label{eq:YqWeakCouplingApprox}
\mathcal{Y}_q(T=0,\VLq\leq\Vq \leq \Vcq)
\approx \left[1-\left(\frac{\Delta_q}{\Delta_q^{(0)}}\frac{\mathbb{V}_{Lq}}{\mathbb{V}_q}\right)^2\right]^{3/2} \, .
\eeqy 
The Landau's velocity~\eqref{eq:Landau-effective-velocity} and the critical velocity~\eqref{eq:Vcq} reduce to the familiar expressions
\beqy\label{eq:Landau-effective-velocity-Approx}
\VLq \approx \frac{\Delta^{(0)}_q}{\hbar k_{Fq}} \, , 
\eeqy 
and 
\beqy \label{eq:Vcq-Approx}
\Vcq\approx \frac{\Euler}{2} \VLq \approx 1.35914 \VLq \, ,
\eeqy 
respectively. 
Here $\Euler \approx 2.71828$ is Euler's number. 

\begin{figure}
\includegraphics[width=10.5cm]{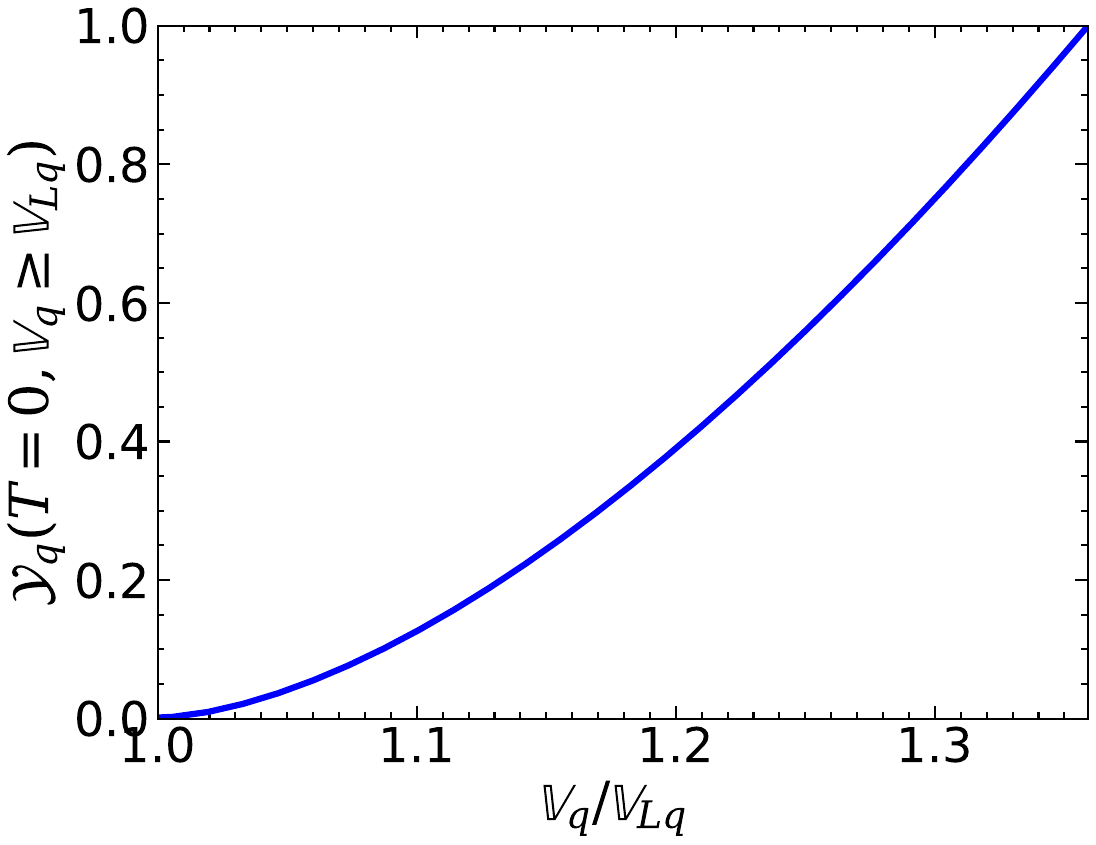}
\caption{Function $\mathcal{Y}_q$ as a function of the effective superfluid velocity $\Vq$ (in units of Landau's velocity $\VLq$) in the gapless superfluid phase of nucleon species $q$ at zero temperature. This function is equal to unity at the critical velocity $\Vcq$. 
}
\label{fig:Yq}
\end{figure}

Equation~\eqref{eq:YqWeakCouplingApprox} is the generalization to 
superfluid mixtures of the expression derived by 
Vollhardt and Maki~\cite{Vollhardt1978} in the context of superfluid 
$^3$He (the function $\mathcal{Y}_q$ was denoted by $\phi$ and was given by 
their Eq.~(18'); their $s$ and $\Delta$ parameters correspond to $\hbar k_{Fq}\mathbb{V}_q$ and $\Delta_q$ respectively). 
As shown in Ref.~\cite{ChamelAllard2021}, the ratio $\Delta_q/\Delta_q^{(0)}$ (at $T=0$) is a universal function of $\mathbb{V}_q/\VLq$ or equivalently of $\mathbb{V}_q/\Vcq$ in this weak-coupling approximation, and it is well fitted by the following formula
\beqy\label{eq:GapInterpolation}
\frac{\Delta_q}{\Delta_q^{(0)}}= 0.5081\sqrt{1-\frac{\mathbb{V}_q}{\mathbb{V}^{(0)}_{cq}}}\left(3.312\frac{\mathbb{V}_q}{\mathbb{V}^{(0)}_{cq}}-3.811\sqrt{\frac{ \mathbb{V}^{(0)}_{cq}}{\mathbb{V}_q}} +5.842\right)\, . 
\eeqy
It follows from Eqs.~\eqref{eq:YqWeakCouplingApprox} and \eqref{eq:GapInterpolation} that $\mathcal{Y}_q$ 
is also a universal function of $\mathbb{V}_q/\Vcq$ at $T=0$, independently of the nucleon species under consideration, the matter composition, and the adopted nuclear energy-density functional. This function is displayed in Fig.~\ref{fig:Yq}.

\subsection{Mass transport and effective superfluid velocity}

We have recently shown that the mass current~\eqref{eq:entrainment1} can be written, in the normal-fluid frame ($\pmb{v_N}=\pmb{0}$) as~\cite{ChamelAllard2020,ChamelAllard2021} :
\beqy
\pmb{\rho_q}\equiv \rho_q \pmb{v_q}=\rho_q \left(1-\mathcal{Y}_q\right)\pmb{\mathbb{V}_q}\, .
\eeqy
From this expression, we can deduce the ``\emph{true}`` velocity $\pmb{v_q}$ (i.e. the velocity with which nucleons are transported) as a non-linear function of the effective superfluid velocity: 
\beqy\label{eq:true-velocity}
\pmb{v_q}=\left(1-\mathcal{Y}_q\right)\pmb{\mathbb{V}_q}\, .
\eeqy

\begin{figure}[h]
\includegraphics[width=10.5cm]{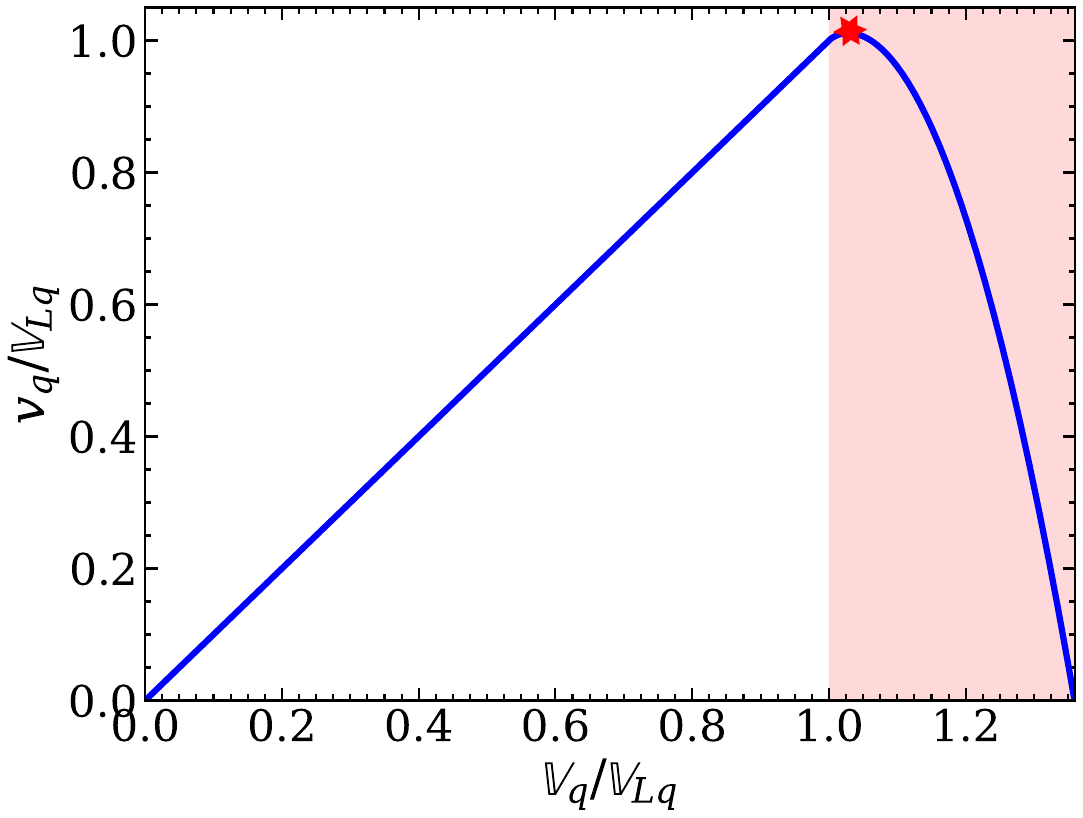}
\caption{True nucleon velocity $v_q$ (associated with transport of nucleon species $q$) as a function of the effective superfluid velocity $\Vq$. Both velocities are expressed in units of Landau's velocity $\VLq$. 
The red star indicates the maximum value $v_q^{\rm (Max)}$ of $v_q$. The shaded area represents the gapless superfluid phase. 
}
\label{fig:HydrodynamicalVelocity}
\end{figure}

In the weak-coupling approximation, it can be easily seen from Eq.~\eqref{eq:YqWeakCouplingApprox} that at zero temperature $v_q/\VLq$ is a universal function of $\Vq/\Vcq$. 
For $\mathbb{V}_q < \mathbb{V}_{Lq}$, we have previously shown that $\mathcal{Y}_q= 0$ so that $\pmb{v_q}$ coincides with $\pmb{\mathbb{V}_q}$. In the gapless superfluid phase characterized by $\VLq\leq \Vq < \mathbb{V}_{cq}^{(0)}$, it can be shown using Eqs.~\eqref{eq:YqWeakCouplingApprox} and \eqref{eq:GapInterpolation} that $1-\mathcal{Y}_q$ decreases with increasing $\Vq$ and so does $v_q$ until the effective superfluid velocity reaches the critical value $\mathbb{V}_{cq}^{(0)}$. At this point and beyond, superfluidity is destroyed, $\mathcal{Y}_q=1$ and $v_q=0$: all nucleons are co-moving with the normal fluid. The variations of $v_q$ with $\Vq$ are plotted in Fig.~\ref{fig:HydrodynamicalVelocity}. 

The maximum value $v^{\rm (Max)}_q$ of the velocity $v_q$ (in the normal frame) is reached for some effective superfluid velocity  $\mathbb{V}^{\rm (Max)}_q$, which is the solution of the following equation: 

\beqy\label{eq:VelocityHydroMax}
\frac{\partial v_q}{\partial \mathbb{V}_{q}}\biggr\vert_{\mathbb{V}_q=\mathbb{V}^{\rm (Max)}_q}=1-\mathcal{Y}_q-\mathbb{V}_{q}\frac{\partial \mathcal{Y}_q}{\partial \mathbb{V}_{q}}\biggr\vert_{\mathbb{V}_q=\mathbb{V}^{\rm (Max)}_q}=0\, .
\eeqy
Solving this equation numerically using Eqs.~\eqref{eq:YqWeakCouplingApprox} and \eqref{eq:GapInterpolation}, yields $v^{\rm (Max)}_q \simeq 1.0111\VLq$ for $\mathbb{V}^{\rm (Max)}_q \simeq 1.0283\VLq$. In the limiting  case of a single constituent (recalling that $\mathbb{V}_{q}=V_q$, as shown in Ref.~\cite{ChamelAllard2021}), our results coincide with those obtained in the BCS theory of electron superconductivity~\cite{Parmenter1962} and those derived for $^3$He superfluidity~\cite{Vollhardt1978}. Our analysis shows that these results remain applicable to mixtures provided the superfluid velocity $V_q$ is replaced by the \emph{effective} superfluid velocity $\Vq$. 
\newpage
\section{Applications to neutron stars}
\label{sec:applications}

For numerical applications to neutron stars, we have adopted the Brussels-Montreal functional BSk24~\cite{goriely2013}, which was constructed from extended Skyrme effective interactions with additional terms that are both density and momentum dependent together with a microscopically deduced pairing interaction. The parameters were precision-fitted to essentially all experimental data on nuclear masses and charge radii while ensuring realistic properties of homogeneous nuclear matter such as the incompressibility, the symmetry energy, effective masses and more importantly for the present study, $^1${\it S}$_0$ pairing gaps. Unified and thermodynamically consistent equations of state spanning all regions of a neutron star (including the mantle of nuclear pastas) have been already calculated for this functional~\cite{pearson2018,shelley2020,pearson2020,pearson2022} allowing for the presence of strong magnetic fields~\cite{mutafchieva2019}. The predictions for the global structure of neutron stars and their tidal deformability are in excellent agreement with astrophysical observations~\cite{perot2019} (higher order gravito-electric and gravito-magnetic tidal deformability parameters have been also computed for this functional in Ref.~\cite{perot2021}).

\subsection{Normal-fluid densities in the outer core: approximate formulas}

In the outer core of a neutron star consisting of $npe\mu$ matter in beta-equilibrium, the normal-fluid fractions $\rho_q^{\rm (N)}/\rho_q$ can deviate from the functions $\mathcal{Y}_q$. However, approximate expressions can be derived remarking that the proton fraction $Y_p=\rho_p/\rho$ does not exceed $10\%$ 
in the region where nucleons are superfluid. 
Expanding in a power series in of $Y_p$, we find to first order
\begin{align}\label{eq:normal-fraction-neutron-approx}
\frac{\rho_n^{\rm (N)}}{\rho_n}&\approx \mneff\frac{ \Yn}{1 + \left(\mneff-1\right) \Yn}+Y_p\frac{\left(\mpeff-1\right) (\Yn-1) \left[\mpeff \Yp (\Yn-1)  + \mneff \Yn(1 - \Yp)\right]}{\mpeff \left[1 + \left(\mneff-1\right) \Yn\right]^2} \, ,
\end{align}

\begin{align}\label{eq:normal-fraction-proton-approx}
\frac{\rho_p^{\rm (N)}}{\rho_p}&\approx \frac{\mpeff (1-\Yn) \Yp + \mneff \Yn \left( \mpeff + \Yp-1\right)}{ \mpeff + \left(\mneff-1\right) \mpeff \Yn} \notag \\
&-Y_p\frac{\left\{\left(\mneff\right)^2  \left( 1 - 2  \mpeff  \right) \Yn + \left(\mpeff\right)^2 \left[1 - \Yn + \mneff (2 \Yn-1)\right] \right\} }{\mneff \left(\mpeff\right)^2 \left[1 + \left(\mneff-1\right)  \Yn\right]^2} \notag \\
&\times  (\Yp-1) \left[\mpeff (\Yn-1) \Yp + \mneff \Yn (1-\Yp)\right] \, , 
\end{align}

\begin{align}\label{eq:normal-fraction-approx}
\frac{\rho_{\rm N}}{\rho}&\approx \frac{\mathcal{Y}_n}{1-\left(\dfrac{m}{m_n^\oplus}-1\right)(\mathcal{Y}_n-1)} \nonumber \\ 
&+Y_p \left\{ 
\left(\frac{m_n^\oplus}{m}-\frac{m_p^\oplus}{m}\right)  (\mathcal{Y}_n-1) \mathcal{Y}_n \left[\frac{m_p^\oplus}{m}\left(1-\mathcal{Y}_n\right) + \frac{m_n^\oplus}{m} \mathcal{Y}_n  \right] \right.\nonumber \\ 
&\left.-\left(\frac{m_p^\oplus}{m}(1-\mathcal{Y}_n)+\frac{m_n^\oplus}{m}\mathcal{Y}_n\right)^2(\mathcal{Y}_n-\mathcal{Y}_p) \right\}
\left\{\frac{m_p^\oplus}{m} \left[1 + \left(\frac{m_n^\oplus}{m}-1\right) \mathcal{Y}_n\right]^2\right\}^{-1}  \, .
\end{align}

In the canonical model of superfluid neutron stars~\cite{andersson2021}, protons are co-moving with leptons 
and quasiparticle excitations; therefore $v_p=0$ in the normal-fluid frame. 
It immediately follows from Eq.~\eqref{eq:true-velocity} that $\mathbb{V}_p=0$ and $\mathcal{Y}_p=0$. The normal-fluid fractions~\eqref{eq:normal-fraction-neutron-approx} and \eqref{eq:normal-fraction-proton-approx} can be further simplified as follows: 

\begin{align}\label{eq:normal-fraction-neutron-approx-oscillations}
\frac{\rho_n^{\rm (N)}}{\rho_n}&\approx \mneff\frac{ \Yn}{1 + \left(\mneff-1\right) \Yn}+Y_p\frac{\left(\mpeff-1\right) (\Yn-1) \mneff \Yn}{\mpeff \left[1 + \left(\mneff-1\right) \Yn\right]^2} \, ,
\end{align}

\begin{align}\label{eq:normal-fraction-proton-approx-oscillations}
\frac{\rho_p^{\rm (N)}}{\rho_p}&\approx \frac{ \mneff \Yn \left( \mpeff  -1\right)}{ \mpeff \left[ 1 + \left(\mneff-1\right) \Yn \right]} \notag \\
&+Y_p\frac{\Yn \left\{\left(\mneff\right)^2 \left( 1- 2 \mpeff \right) \Yn   + \left(\mpeff\right)^2 \left[1 - \Yn + \mneff (2 \Yn-1)\right] \right\} }{ \left( \mpeff\right)^2 \left[1 + \left(\mneff-1\right) \Yn\right]^2}    \, . 
\end{align}

\subsection{Normal-fluid densities in the outer core: numerical results}

\begin{figure}
\includegraphics[width=10.5cm]{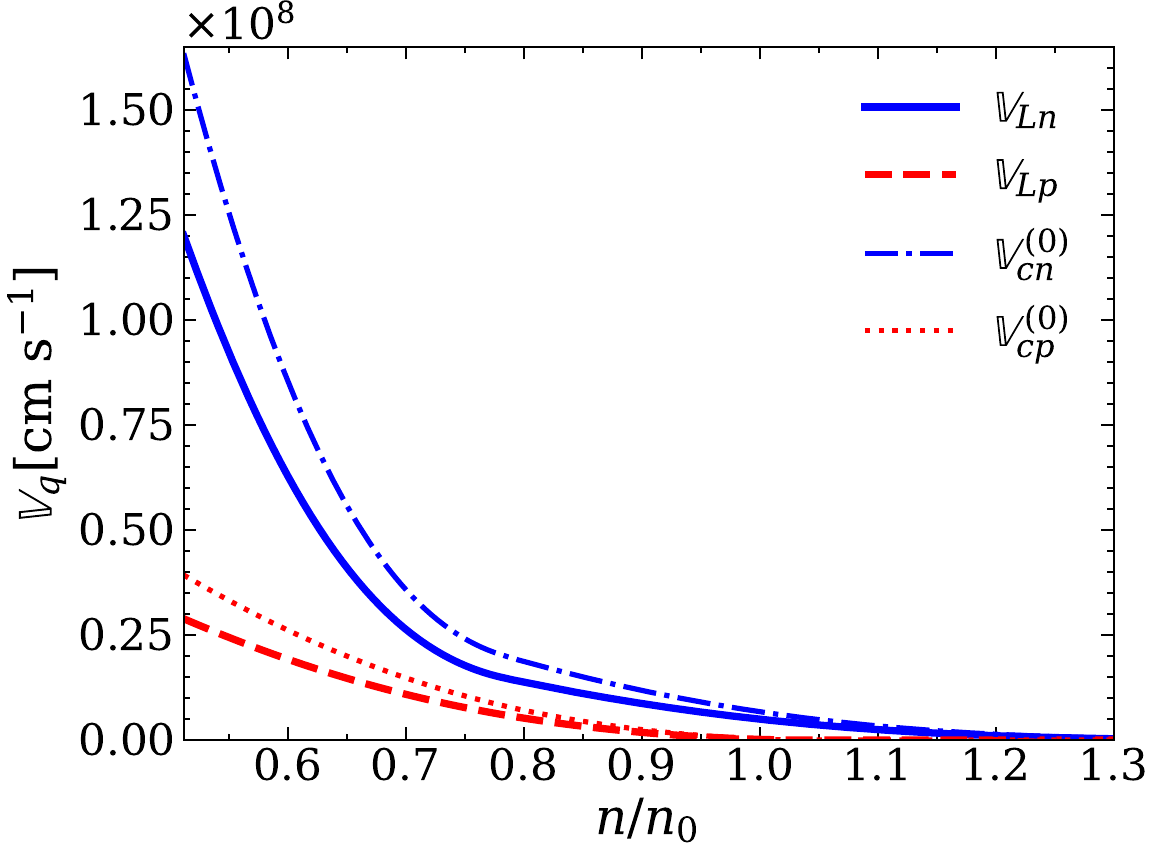}
\caption{Landau's and critical effective superfluid velocities for neutrons and protons (in units of cm~s$^{-1}$) as a function of the baryon number density $n$ (in units of the saturation density $n_0$) for $npe\mu$ matter in beta-equilibrium in the outer core of a cold neutron star. 
}
\label{fig:VLqCore}
\end{figure}

We have computed the normal-fluid fractions using Eqs.~\eqref{eq:NormalDensity1}-\eqref{eq:normal-fluid-fraction} and Eqs.~\eqref{eq:EntrainmentMatrix}-\eqref{eq:Theta-def} together with Eq.~\eqref{eq:YqTZero} where $\Dbar$ and  $\mubar$ are obtained from the numerical solutions of Eqs.~\eqref{eq:GapZeroT} and \eqref{eq:MuZeroT}. Landau's velocities $\VLq$ are calculated using Eq.~\eqref{eq:Landau-effective-velocity} and the critical velocities $\Vcq$ are determined by solving Eqs.~\eqref{eq:Vcq}-\eqref{eq:MuZeroTVc}. 
These velocities are plotted in Fig.~\ref{fig:VLqCore}. 

\begin{figure}
\includegraphics[width=10.5cm]{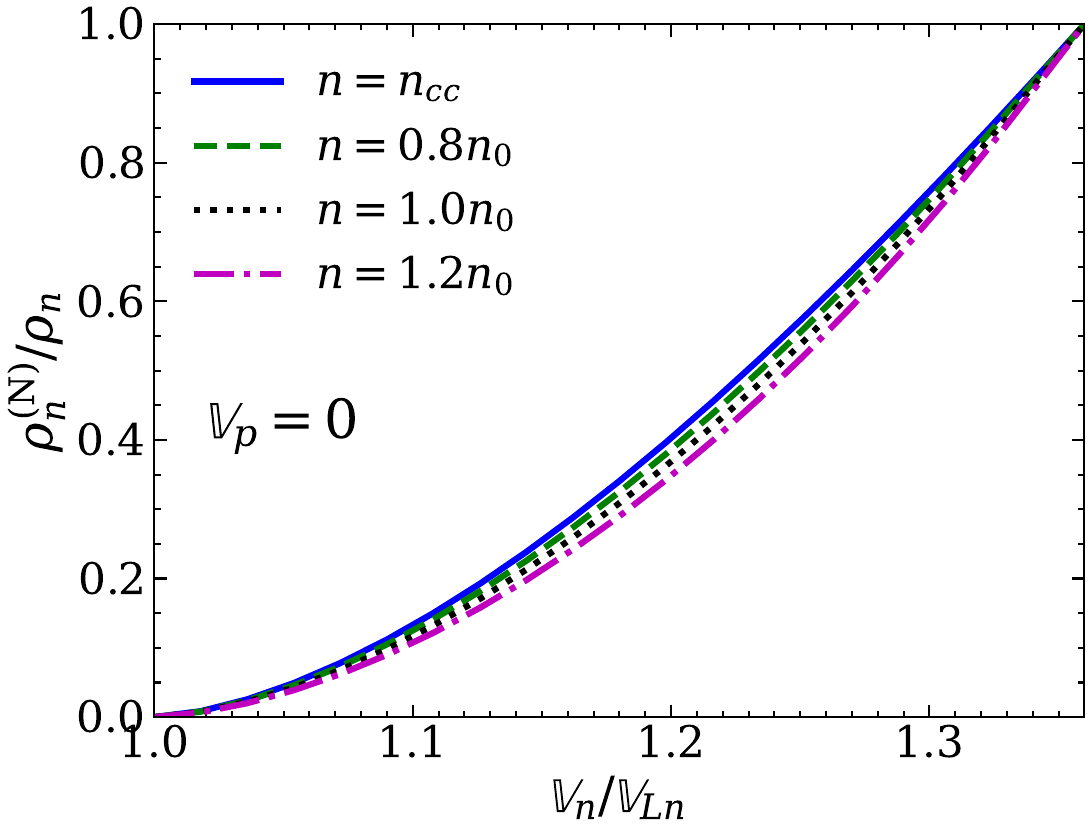}
\caption{Normal-fluid fraction associated with neutron quasiparticles as a function of the neutron effective  superfluid velocity $\mathbb{V}_n$ (in units of Landau's velocity $\mathbb{V}_{Ln}$) for $npe\mu$ matter in $\beta$-equilibrium in the outer core of a cold neutron star and for effective proton superfluid velocity $\mathbb{V}_p=0$. 
 }
\label{fig:NormalFractionNeutronBetaBSk24}
\end{figure}

The normal-fluid fraction $\rho_n^{\rm (N)}/\rho_n$ associated with neutrons in the gapless phase is shown in Fig.~\ref{fig:NormalFractionNeutronBetaBSk24} for baryon densities ranging from the crust-core transition density $n_{\rm cc}\simeq 0.081~{\rm fm}^{-3}\approx n_0/2$ ($n_0\simeq 0.1578~{\rm fm}^{-3}$ is the saturation density of symmetric nuclear matter) up to the point where neutron superfluidity is destroyed at density $n\approx 1.3n_0$. 
As expected from the approximate formula~\eqref{eq:normal-fraction-neutron-approx-oscillations}, $\rho_n^{\rm (N)}/\rho_n$ 
is weakly dependent on the density, and is mainly determined by $\mathcal{Y}_n$. The normal-fluid fraction, which vanishes for $\mathbb{V}_n\leq \mathbb{V}_{Ln}$, increases monotonically with the neutron effective superfluid velocity as more and more quasiparticles are being excited until the maximum is reached at the critical velocity $\mathbb{V}_{cn}^{(0)}$. 

\begin{figure}
\includegraphics[width=10.5cm]{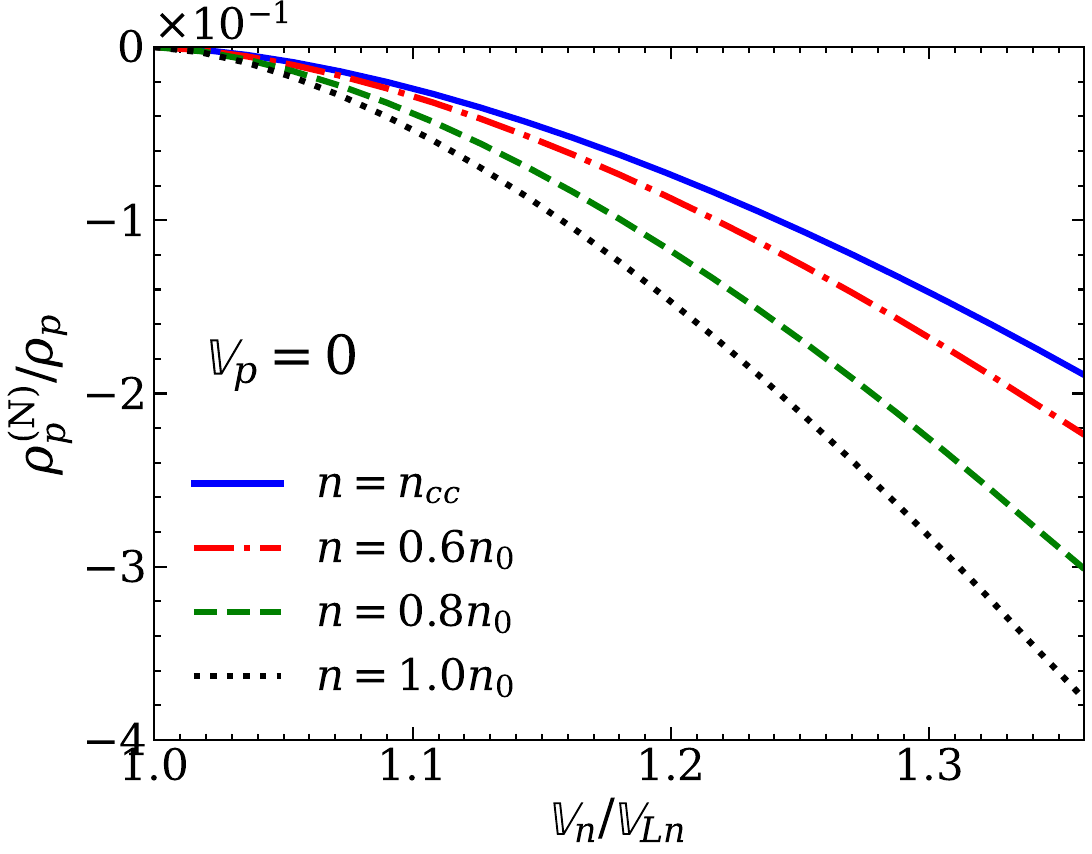}
\caption{Same as Fig.~\ref{fig:NormalFractionNeutronBetaBSk24} but for protons.}
\label{fig:NormalFractionProtonBetaBSk24}
\end{figure}

As shown in Fig.~\ref{fig:NormalFractionProtonBetaBSk24}, the normal-fluid fraction $\rho_p^{\rm (N)}/\rho_p$ associated with protons is found to be negative. This result may seem surprising at first sight, but one has to remember that the normal and superfluid densities are current-current response functions (see, e.g., Ref.~\cite{Baym1968}), and therefore are not the densities of anything, as stressed by Feynman~\cite{Feynman1955} in the context of superfluid helium. Landau himself emphasized in his seminal paper about the two-fluid model~\cite{Landau1941} that ``there is no division of the real particles of the liquid into `superfluid' and `normal' ones''. The negative values of $\rho_p^{\rm (N)}$ can be understood from the approximate formula~\eqref{eq:normal-fraction-proton-approx-oscillations}, recalling that $Y_p\ll 1$ and noticing that $m_p^\oplus<m_p$.

\begin{figure}
\includegraphics[width=10.5cm]{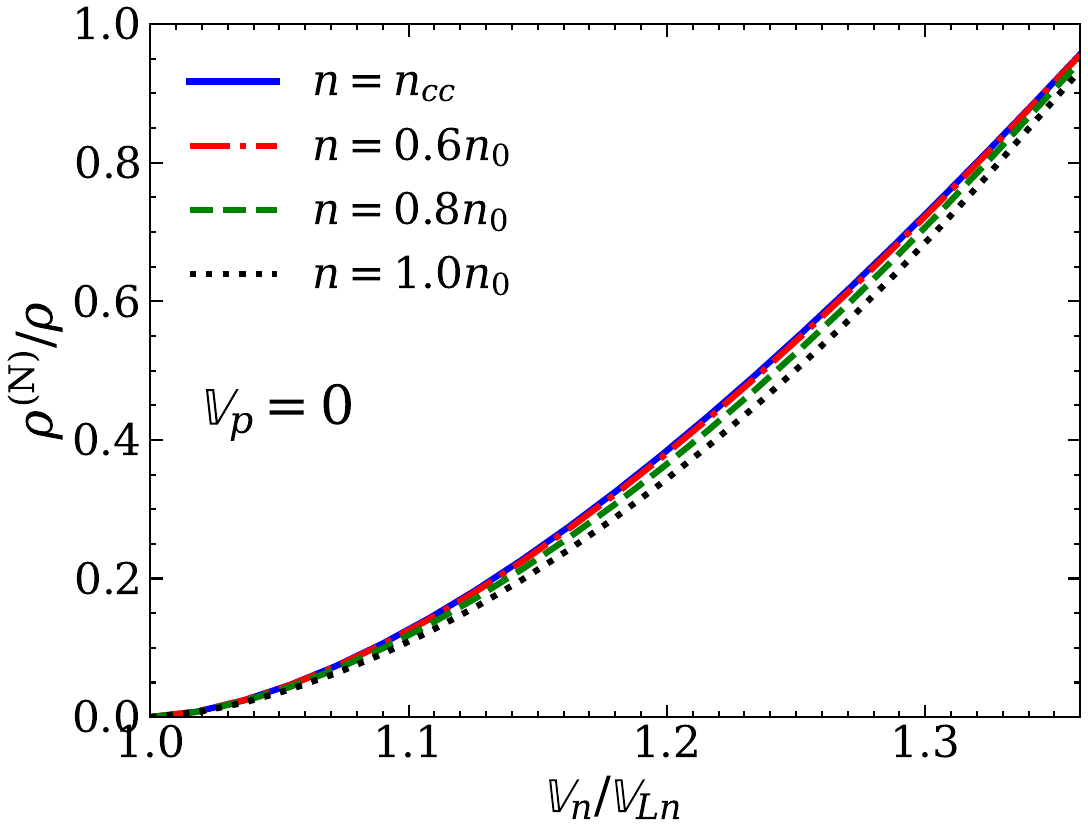}
\caption{Same as Fig.~\ref{fig:NormalFractionNeutronBetaBSk24} but for the total normal-fluid fraction.}
\label{fig:NormalFractionTotalBetaBSk24}
\end{figure}

The total mass density $\rho_{\rm N}$ carried by the quasiparticle excitations and defined by Eq.~\eqref{eq:normal-fluid-fraction} is plotted in Fig.~\ref{fig:NormalFractionTotalBetaBSk24}. 
Note that for neutron superfluid velocities $\Vn>\mathbb{V}_{cn}^{(0)}$, the total normal-fluid fraction remains lower than one and decreases with increasing density. Although neutrons are in the normal phase ($\rho_n^{(\rm N)}=\rho_n$), protons remain superfluid since $\mathbb{V}_p=0$. Setting $\mathcal{Y}_n=1$ in Eq.~\eqref{eq:normal-fraction-proton-approx-oscillations}, the total normal-fluid fraction  $\rho_{\rm N}/\rho = (1-Y_p)\rho_n^{\rm (N)}/\rho_n  + Y_p\rho_p^{\rm (N)}/\rho_p $
is approximately given by
\begin{align}
\frac{\rho_{\rm N}}{\rho} \approx 1 - Y_p  \frac{m}{m_p^\oplus} \, .
\end{align}

\begin{figure}[h]
    \centering
    \includegraphics[width=0.7\textwidth]{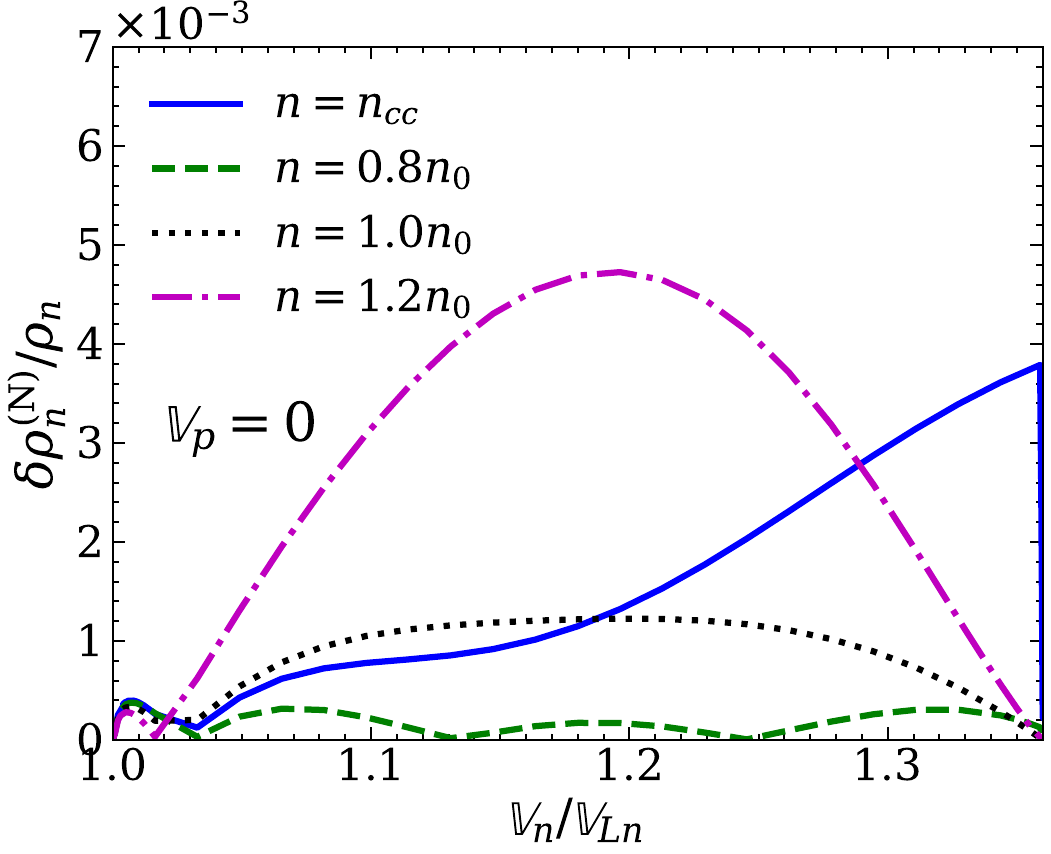}
    \caption{Deviations between the normal-fluid fraction for neutrons plotted in Fig.~\ref{fig:NormalFractionNeutronBetaBSk24} and the approximate formula~\eqref{eq:normal-fraction-neutron-approx-oscillations} combined with Eqs.~\eqref{eq:YqWeakCouplingApprox} and~\eqref{eq:GapInterpolation}, as a function of the neutron effective superfluid velocity $\Vn$ (in units of Landau's velocity $\VLn$).}
    \label{fig:EAFracNBetaBSk24OscillYApprox}
\end{figure}

\begin{figure}[h]
    \centering
    \includegraphics[width=0.7\textwidth]{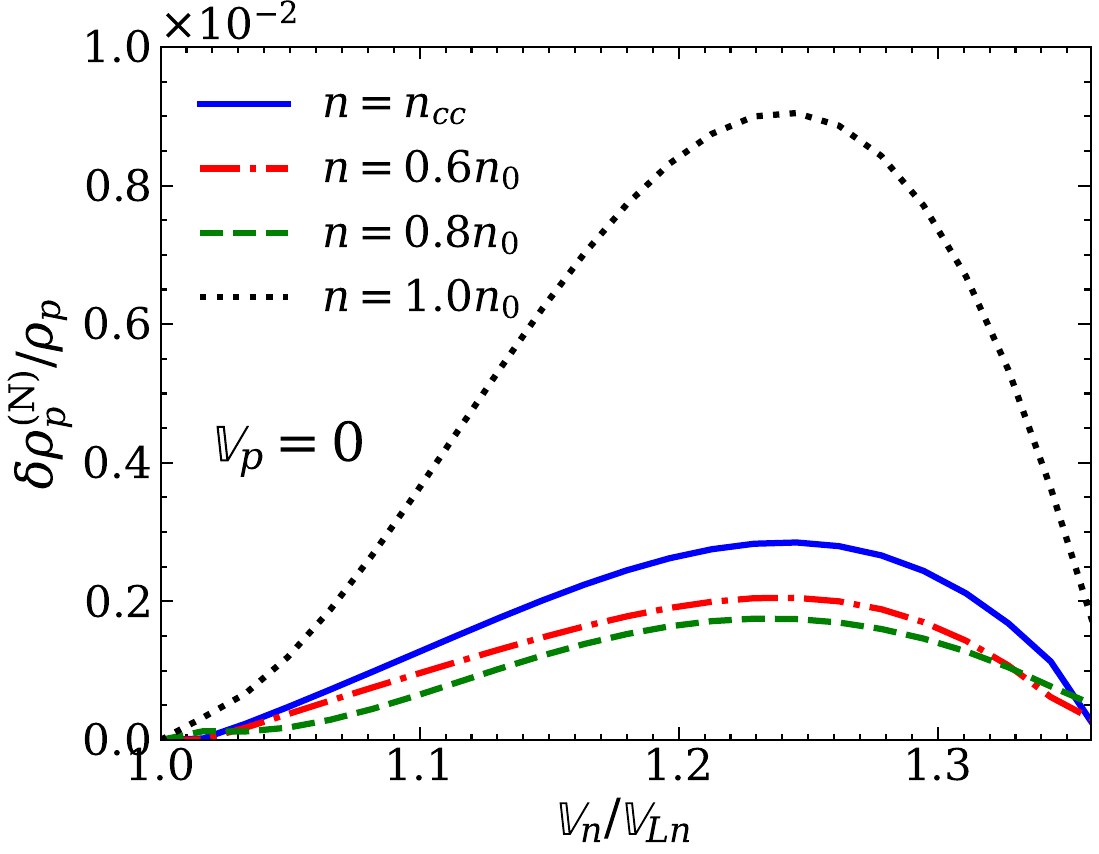}
    \caption{Same as Fig.~\ref{fig:EAFracNBetaBSk24OscillYApprox} for protons.}
    \label{fig:EAFracPBetaBSk24OscillYApprox}
\end{figure}

For comparison, we have estimated the normal-fluid densities using the expansions~\eqref{eq:normal-fraction-neutron-approx-oscillations} and \eqref{eq:normal-fraction-proton-approx-oscillations} together with the approximate universal formulas for $\mathcal{Y}_q$ and $\Delta_q/\Delta_q^{(0)}$ given by Eqs~\eqref{eq:YqWeakCouplingApprox} and~\eqref{eq:GapInterpolation} respectively. 
We have evaluated the dimensionless ratio $\Vn/\VLn$ using the 
exact expression~\eqref{eq:Landau-effective-velocity} for Landau's velocity rather than the approximate formula for a more direct comparison. However, we have found that the approximate formula~\eqref{eq:Landau-effective-velocity-Approx} for neutrons deviates from the exact result by $\sim 0.06$\%  at the crust-core transition and the error decreases with increasing density. Let us recall that Eq.~\eqref{eq:GapInterpolation} tacitly assumes that the critical velocity is given by Eq.~\eqref{eq:Vcq-Approx}. The errors on $\mathbb{V}_{cn}^{(0)}$ amount to $\sim 0.1$\% at most. The ratio $\mathbb{V}_{cn}^{(0)}/\VLn$ is given by $\exp(1)/2$ within $\sim 0.088$\%. 
As shown in Figs.~\ref{fig:EAFracNBetaBSk24OscillYApprox} and \ref{fig:EAFracPBetaBSk24OscillYApprox}, the deviations $\delta \rho_q^{({\rm N})}/\rho_q$ between the exact and the approximate results do not exceed $5\times 10^{-3}$ for neutrons and $10^{-2}$ for protons. 

\newpage
\section{Conclusions}
\label{sec:conclusions}

With the possibility of gapless nuclear superfluidity, the hydrodynamics of cold neutron stars may 
be much more complicated than previously thought. In particular, a normal fluid consisting of quasiparticle 
excitations could be present inside the star even at zero temperature if the effective superfluid 
velocity $\Vq$ exceeds Landau's velocity $\VLq$ while lying below the critical velocity $\Vcq$. 
To show this, we have calculated explicitly the normal-fluid densities $\rho_n^{\rm(N)}$ and 
$\rho_p^{\rm(N)}$ for arbitrary neutron-proton superfluid mixtures within the time-dependent 
nuclear energy-density functional theory. We have thus confirmed that these densities become 
nonzero in the gapless phase and their dependencies on $\Vq$ are entirely contained in the 
functions $\mathcal{Y}_q$, for which we have obtained an exact analytical expression, namely 
Eq.~\eqref{eq:YqTZero}, in terms of the reduced chemical potential $\mubar$ and the pairing 
parameter $\Dbar$. In the weak-coupling approximation, $\mathcal{Y}_q$ becomes a universal 
function of the dimensionless ratio $\Vq/\VLq$, independently of the composition, the total 
mass density $\rho$ and the adopted nuclear energy-density functional. Whereas in the subcritical 
regime $\Vq$ coincides with the ``true'' velocity $v_q$ with which nucleons are actually transported, 
these two kinds of velocities differ in the gapless phase. These velocities are also different 
from the superfluid velocity $V_q$, which is fundamentally the momentum per unit mass carried 
by the superfluid, as discussed in Section 2.4 of Ref.~\cite{ChamelAllard2021}. 

For applications to neutron stars, we have derived approximate analytical expressions of the 
normal-fluid fractions to first order in the proton fraction $Y_p=\rho_p/\rho$. We have also 
evaluated $\rho_n^{\rm(N)}/\rho_n$, $\rho_p^{\rm(N)}/\rho_p$, and $\rho_{\rm N}/\rho$ numerically 
by solving the full TDHFB equations (including the equations for $\mubar$ and $\Dbar$) using the 
functional BSk24~\cite{goriely2013} for which the composition of a neutron star has been already 
calculated in all regions~\cite{pearson2018}. 

Together with the unified equation of state and the superfluid properties published in 
Refs.~\cite{pearson2018,ChamelAllard2021} and available in the CompOSE database~\cite{typel2021}, the results presented here provide the microscopic 
inputs for modelling the dynamics of cold neutron stars. Our study suggests that the neutron 
superfluid reservoir in the outer core may be substantially reduced in the gapless phase, 
thus challenging even further the interpretation of pulsar glitches~\cite{andersson2012,chamel2013}.
The question that arises is whether 
the neutron and proton superfluids can reach the gapless phase. Already in 1998, Sedrakian and 
Cordes~\cite{sedrakian1998} considered the possibility that neutron superfluid velocities induced 
by the rotational evolution of pulsars could exceed Landau's velocity~\eqref{eq:Landau-effective-velocity-Approx}. But they implicitly assumed that superfluidity would be destroyed. 
%Let us examine this issue more closely. 
In the two-fluid picture of superfluid neutron stars, protons and leptons are co-moving 
with the normal fluid of quasiparticle excitations so that the proton effective superfluid 
velocity $\mathbb{V}_p=0$ in the normal-fluid rest frame. Therefore, protons are not expected 
to be in the gapless phase. The normal fluid slows down due to electromagnetic braking 
but can also accelerate due to accretion from a stellar companion in a binary system. 
On the other hand, the evolution of the neutron superfluid is dictated by the dynamics 
of neutron quantized vortices. In a fixed external frame in which the star is rotating 
with the velocity $\pmb{v_N}$, the neutron superfluid velocity $\pmb{V^\prime_n}=\pmb{V_n}+\pmb{v_N}$ 
remains unchanged for as long as neutron vortices are pinned. Indeed, the quantization of 
the neutron superflow imposes 
\beqy 
\oint_{\mathcal{C}} \pmb{V^\prime_n}\cdot \pmb{{\rm d}\ell}=N\frac{h}{2m_n}
\eeqy 
along any contour $\mathcal{C}$ enclosing $N$ vortices. In the normal-fluid frame, 
$\Vn$, which is roughly equal to $V_n$ (see Section 3.6 of Ref.~\cite{ChamelAllard2021}), 
therefore increases with time. The lag between the neutron superfluid and the rest of the 
star induces a Magnus force acting on the vortices. The maximum neutron superfluid velocity 
is therefore limited by the critical lag $V_{\rm cr}$ for which the Magnus force equals 
the pinning force.

The pinning force remains very uncertain and even its nature (attractive 
or repulsive) is a matter of debate. Estimates differ by several orders of 
magnitude. In their seminal paper, Anderson and Itoh~\cite{anderson1975} 
estimated the force per unit length as $f_p\sim 10^{20}p$~dyn/cm, where 
$p\sim 0.01-0.1$ is the pinning probability. Two years later, Alpar~\cite{Alpar1977} 
refined this estimate by taking into account realistic pairing gaps and crust 
profiles and found that the force varies depending on the layer and is of 
order $10^{18}$~dyn/cm. The pinning force depends on the quantum structure 
of a vortex as well as the interactions between the constituent neutrons and 
the dense medium. The first quantum calculations were carried out by Avogadro 
and collaborators~\cite{avogadro2007}, who determined the energy gain due to 
the pinning of a vortex segment on a cluster in a Wigner-Seitz cell (see 
Ref.~\cite{klausner2023} for more recent calculations). However, 
calculations of energy differences can be very delicate and sensitive to boundary 
conditions. The results obtained with this approach were questioned by Pizzochero 
and collaborators~\cite{pizzochero2008}, who had calculated earlier the energies 
using a semiclassical treatment~\cite{pizzochero1997,donati2003,donati2006} (but see 
also Ref.~\cite{avogadro2008}). 
Alternatively, a more 
reliable approach is to calculate the force dynamically, as proposed in Ref.~\cite{bulgac2013}. 
Fully microscopic dynamical simulations of the interactions between a single 
vortex and a nucleus have only been recently tackled, see, e.g., Ref.~\cite{wlazlowski2016}. 
Systematic calculations still remain to be performed. On top of that, another 
challenge is to understand the large-scale dynamics of each vortex in the presence 
of a lattice of pinning sites~\cite{seveso2016,antonelli2020}. Depending on how strong 
the vortex tension is and how nuclei are 
distributed, a vortex line could bend so as to maximize the number of pinned nuclei 
resulting in much stronger mean pinning forces than simple estimates considering a 
straight vortex, see e.g. Ref.~\cite{Link2022} and references therein.  
More importantly, neutron vortices could also be pinned to proton fluxoids in the deep crust~\cite{ZhaoWen2021} (where some protons can be unbound) and in the outer core~\cite{muslimov1985vortex,srinivasan1990novel,sauls1989superfluidity,ruderman1998neutron,alpar2017}, assuming protons form a type II superconductor~\cite{baym1969} (but see also Refs.~\cite{charbonneau2007,alford2008,wood2022}). Recent timing observations of 
the Crab and Vela pulsars during the rise of a glitch bring some support to these  
additional pinning sites~\cite{sourie2020}. 

The critical lag for vortex unpinning is even more uncertain and model 
dependent. Indeed, the ``local" angular velocities are determined by averaging 
over a matter element containing a large collection of individual vortices. 
According to the snow plow model of Ref.~\cite{pizzochero2011} in which straight 
parallel vortices can pin to the crust, the maximum superfluid velocity 
can be estimated as  $V_{\rm cr}\sim 10^7(f_p/10^{18}~{\rm dyn/cm})$~cm/s. 
For comparison, $\VLn$ takes its maximum value at the crust-core transition 
but rapidly decreases with increasing density, as can be seen in Fig.~\ref{fig:VLqCore}. 
At density $\sim 0.65 n_0$, $\VLn$ drops by about an order of magnitude down 
to $\sim 10^7$~cm/s. 
Therefore, the existence of gapless superfluidity in neutron-star cores is not 
implausible and its astrophysical implications should be further studied. 

\begin{acknowledgments}
This work was financially  supported by the Fonds de la Recherche Scientifique (Belgium) under 
Grant No. PDR T.004320. The authors thank Prof Armen Sedrakian for discussions.
\end{acknowledgments}

\appendix
\section{Normal-fluid fraction in the absence of currents}

In the limit of small currents such that $\beta\hbar \pmb{k}\cdot \pmb{\mathbb{V}_q} \rightarrow 0$, the function~\eqref{eq:YqFunction} reduces to 
\beqy
\mathcal{Y}_q(T)=\dfrac{\beta \hbar^2}{2m_q^\oplus n_q} \dfrac{1}{V}\sum_{\pmb{k}} k^2\cos^2\theta_{\pmb{k}} \sech^{2}\left( \frac{\beta}{2} \sqrt{\varepsilon^{(q)2}_{\pmb{k}} +  \Delta_q^2}\right)\, ,
\eeqy 
where $\theta_{\pmb{k}}$ denotes the angle between $\pmb{k}$ and $\pmb{\mathbb{V}_q}$. 
Taking the continuum limit~\eqref{eq:continuum} and integrating over solid  angle in $\pmb{k}$-space yield
\beqy\label{eq:Yq-static}
\mathcal{Y}_q(T)=\dfrac{1}{6}\dfrac{\beta \hbar^2}{m_q^\oplus n_q} \int_{-\mu_q}^{+\infty} {\rm d}\varepsilon\, \mathcal{D}_q(\varepsilon)  k^2 \sech^{2}\left( \frac{\beta}{2} \sqrt{\varepsilon^2 +  \Delta_q^2}\right)\, ,
\eeqy 
recalling that $k$ is related to $\varepsilon$ through Eq.~\eqref{eq:varepsilon}. 

In the spirit of Leggett's theory of superfluid Fermi liquids~\cite{leggett1965} assuming that $T_{cq}\ll T_{Fq}$, 
we introduce the following approximations to simplify Eq.~\eqref{eq:Yq-static}: 
\begin{itemize}
    \item the density of single-particle states per spin $\mathcal{D}(\varepsilon)$ is replaced by its value on the Fermi surface, $\mathcal{D}(\varepsilon)\approx m_q^\oplus k_{Fq}/(2\pi^2 \hbar^2)$ using Eq.~\eqref{eq:DoS}; 
    \item instead of solving Eq.~\eqref{eq:DensityHomogeneous}, the reduced chemical potential is approximated by the Fermi energy, $\mu_q\approx \hbar^2 k_{Fq}^2/(2 m_q^\oplus)$ with $k_{Fq}=(3\pi^2 n_q)^{1/3}$; 
    \item the factor $k^2$ is replaced by $k_{Fq}^2$. 
\end{itemize} 
With these approximations, the function $\mathcal{Y}_q$ becomes 
\beqy
\mathcal{Y}_q(T)\approx \dfrac{\beta }{4}  \int_{-\mu_q}^{+\infty} {\rm d}\varepsilon\,  \sech^{2}\left( \frac{\beta}{2} \sqrt{\varepsilon^2 +  \Delta_q^2}\right)\, .
\eeqy 
Introducing the variable $y\equiv \varepsilon/(k_B T_{cq})$, we find
\beqy
\mathcal{Y}_q(T) =  \dfrac{T_{cq} }{4 T}  \int_{-T_{Fq}/T_{cq}}^{+\infty} {\rm d}y \,  \sech^{2}\left[ \frac{T_{cq}}{2 T} \sqrt{y^2 +  \left(\dfrac{\Delta_q}{k_B T_{cq}}\right)^2}\right]\, .
\eeqy 
Assuming $T_{cq}\ll T_{Fq}$ and noticing that the integrand falls rapidly to zero for increasing $\vert y\vert$, we can thus replace the lower bound of the integral by $-\infty$. Since the integrand remains invariant under the change of $y$ by $-y$, we finally obtain
\beqy
\mathcal{Y}_q(T) \approx \dfrac{T_{cq} }{2 T}  \int_{0}^{+\infty} {\rm d}y \,  \sech^{2}\left[ \frac{T_{cq}}{2 T} \sqrt{y^2 +  \left(\dfrac{\Delta_q}{k_B T_{cq}}\right)^2}\right]\, .
\eeqy 
This expression coincides with the Yosida function~\cite{yosida1958} denoted by 
$f(T)$ in Ref.~\cite{leggett1965}. 
In the limit of a static superfluid, it can thus be seen that the normal-fluid fraction~\eqref{eq:normal-fraction-approx2}
reduces to Eq.~(72) of Ref.~\cite{leggett1965}.

\end{document}